\input{aipcheck}

\documentclass[
    ,final            
  ]
  {aipproc}

\layoutstyle{6x9}

\begin{document}

\title{The big picture of AGN feedback: Black hole accretion and galaxy evolution in 
multiwavelength surveys }

\classification{98.54.Cm, 98.54.Gr, 98.62.Ai}
\keywords      {galaxies:active, galaxies:evolution, X-rays:galaxies}

\author{Ryan C.\ Hickox}{
  address={Harvard-Smithsonian Center for Astrophysics, 60 Garden Street, Cambridge, MA  02138}
}

\begin{abstract}
 
Large extragalactic surveys allow us to trace, in a statistical sense,
how supermassive black holes, their host galaxies, and their dark
matter halos evolve together over cosmic time, and so explore the
consequences of AGN feedback on galaxy evolution.  Recent studies have
found significant links between the accretion states of black holes
and galaxy stellar populations, local environments, and obscuration by gas
and dust.  This article describes some recent results and shows how
such studies may provide new constraints on models of the co-evolution
of galaxies and their central SMBHs.  Finally, I discuss observational
prospects for the proposed {\em Wide-Field X-ray Telescope} mission.

\end{abstract}

\maketitle


\section{Introduction}

Recently, a number of galaxy evolution models have posited that 
feedback from active galactic nuclei (AGN) can play a key role in
producing the observed evolution of galaxies and clusters (see
\citep{hickox_r_catt09bhgal} for a review).  
Feedback is observed most directly in clusters and groups, where AGN
outflows interact strongly with hot X-ray emitting gas
\citep{hickox_r_mcna07araa}. However, for galaxies in rarer environments or higher redshifts, the effects of AGN feedback are most apparent in the evolution of galaxies' stars and cold gas.

Large surveys comprising many thousands of galaxies and active
galactic nuclei (AGN) have enabled precision measurements of
luminosity functions, clustering, and evolution with redshift of both
galaxies \citep[e.g.,][]{hickox_r_fabe07lfunc,hickox_r_brow08halo} and accreting
supermassive black holes (SMBHs; \citep[e.g.,][]{hickox_r_ueda03,
hickox_r_croo05}). Constrained by these observables, recent models have invoked
different modes of feedback: radiatively-dominated (``quasar'') modes,
in which luminous quasars blow gas out of galaxies and terminate
episodes of star formation, and kinetically-dominated (``radio'')
modes, in which lower-efficiency accretion produces outflows that
prevent further cooling to form stars.

Models with these prescriptions have successfully reproduced, in broad
terms, the luminosities, colors, and clustering of galaxies (derived
mainly from optical and IR surveys)
\citep[e.g.,][]{hickox_r_bowe06gal, hickox_r_crot06} as well as those
of AGN (derived primarily from X-ray, optical, and radio data)
\citep[e.g.,][]{hickox_r_hopk08frame1, hickox_r_merl08agnsynth}.
However, in constraining models the galaxy and AGN observables are
often treated {\em separately}, while the very nature of feedback
implies that AGN and galaxies evolve {\em together}.  Successful
feedback models must therefore predict not only the observed
distribution of AGN and galaxy properties, but AGN in different
accretion states must be found in the appropriate host galaxies and
environments.

\begin{figure}
  \includegraphics[width=.42\textwidth]{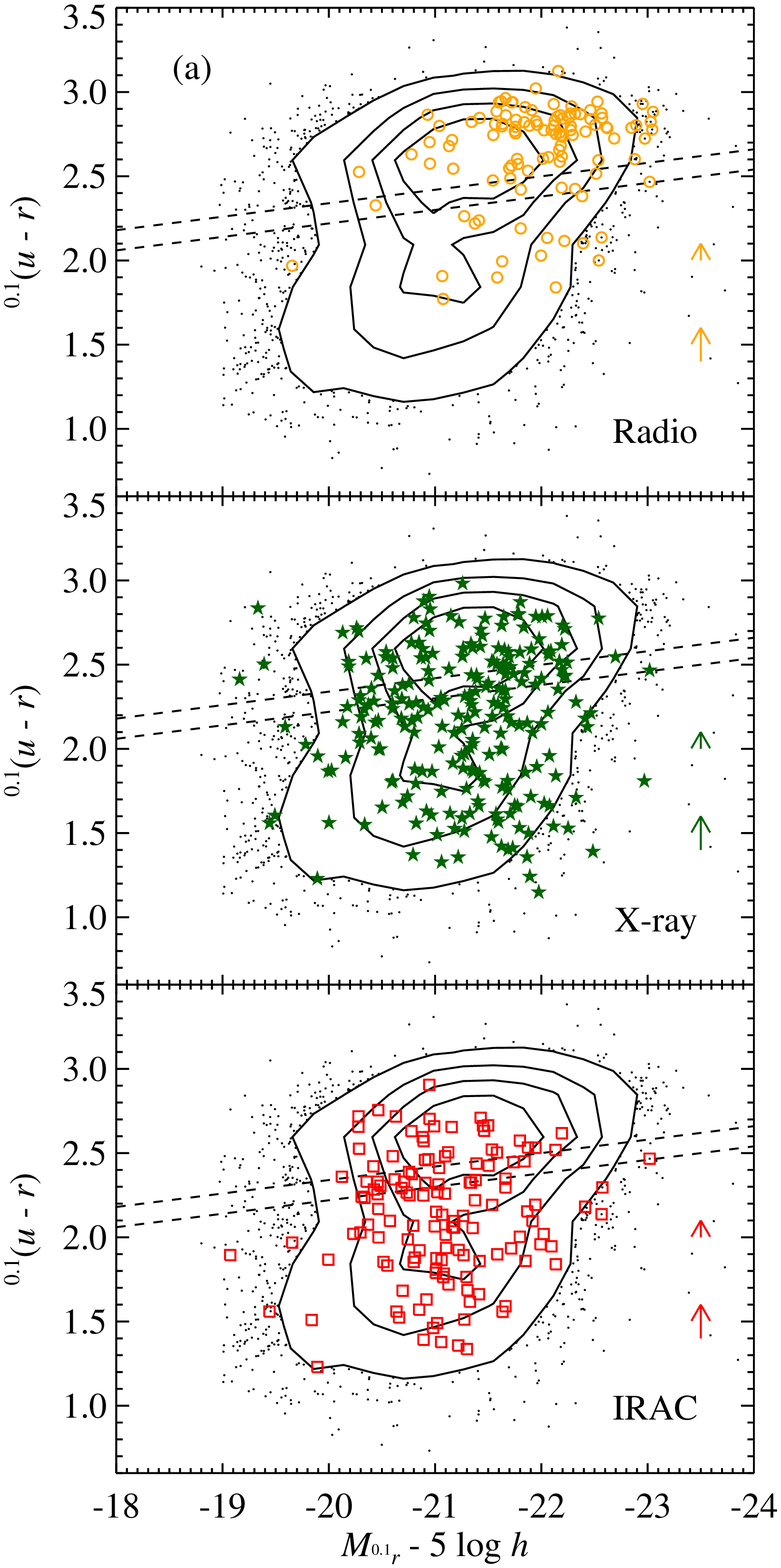}

  \includegraphics[width=.50\textwidth]{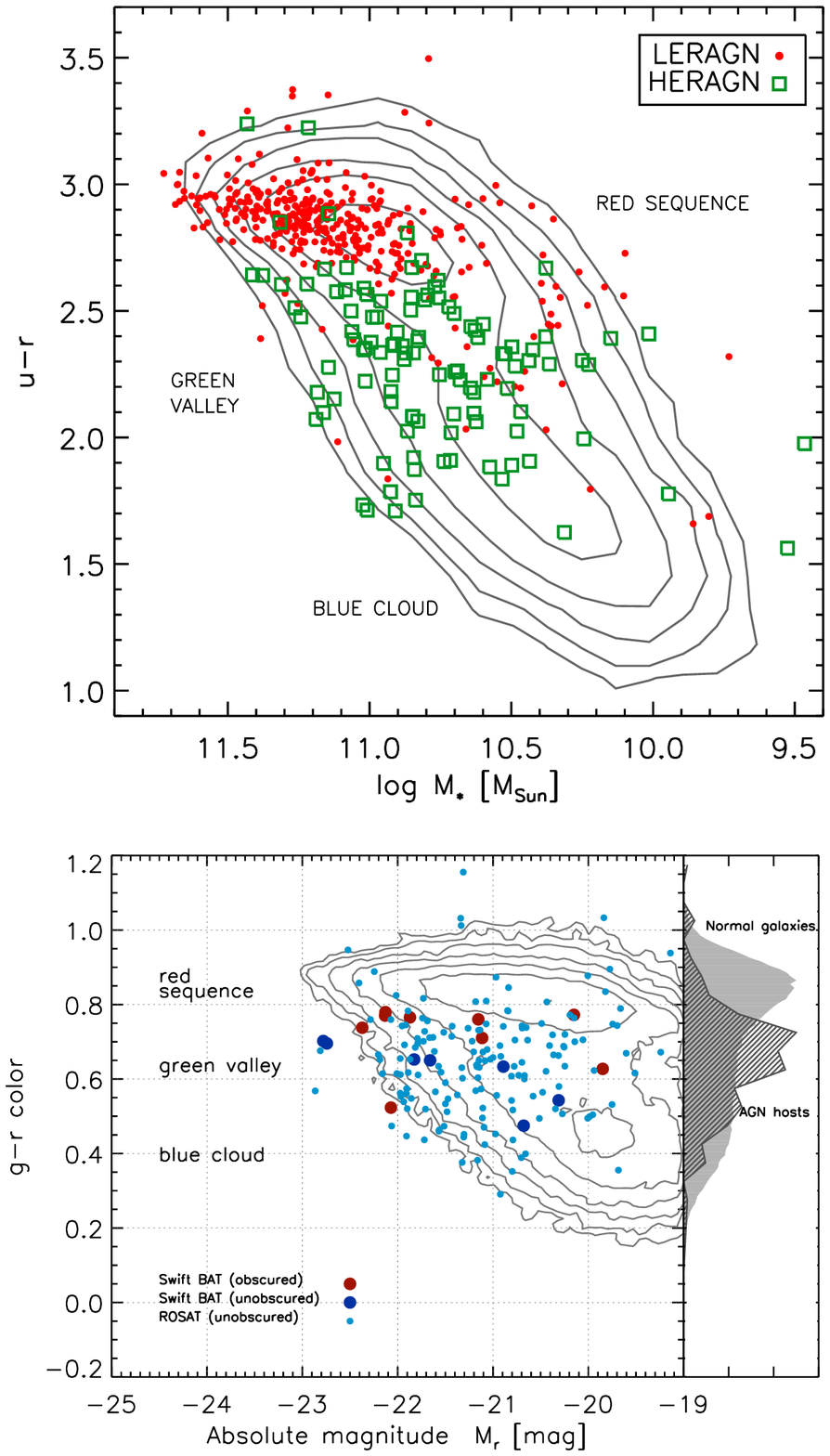}

  \caption{Colors and luminosities for AGN host galaxies, for AGN
  detected in the radio, X-ray and IR bands \citep{hickox_r_hick09corr} ({\em
  left}), for radio AGN with high- and low-excitation emission line
  spectra \citep{hickox_r_smol09radio} ({\em top right}), and for X-ray AGN
  \citep{hickox_r_scha09agn} ({\em bottom right}).  AGN in radiatively
  efficient states are typically found in blue cloud and ``green
  valley'' galaxies, while low-Eddington AGN populate
    the red sequence.  Figures reproduced with permission of AAS.
\label{fig:colmag}}
\end{figure}

\section{AGN and galaxy evolution in multiwavelength surveys}

Large multiwavelength surveys allow us to test these feedback
scenarios directly, by observing galaxies and AGN within the same
statistical samples.  For example, \citet{hickox_r_hick09corr} used data from
the 9 deg$^2$ Bo\"{o}tes multiwavelength survey to study the AGN and
galaxy population at $z\sim 0.5$.  Selecting AGN in the radio, X-rays,
and infrared (IR), Hickox et al.\ found that AGN in different
accretion states are found in markedly different host galaxies and
large-scale environments (Fig.~\ref{fig:colmag}, {\em left}).  Radio
AGN, which primarily have low Eddington ratios ($\lambda<10^{-3}$),
are found in strongly clustered, luminous red galaxies, while IR AGN
have high $\lambda>10^{-2}$ and are found in weakly clustered, less
luminous and bluer galaxies.  X-ray AGN have intermediate to high
$\lambda$ and reside preferentially in ``green'' galaxies with
clustering typical of the full galaxy population.  \citet{hickox_r_smol09radio}
obtained similar results for a radio-selected sample in SDSS, with
high-excitation (high-$\lambda$) radio AGN residing in ``green''
galaxies while the low-excitation (low-$\lambda$) objects lie on the
red sequence (Fig.~\ref{fig:colmag}, {\em top right}).
Correspondingly,
\citet{hickox_r_kauf09modes} found that in SDSS spectroscopic 
samples, the Eddington ratio distribution peaks at $\lambda\sim$ a few
percent for galaxies with young stellar populations, but for older
galaxies the distribution rises to very low Eddington ratios.  These and a
number of other recent results
\citep[e.g.,][]{hickox_r_bund08quench, hickox_r_silv08host} suggest that black hole accretion broadly tracks the history of star formation in galaxies, with smaller, younger systems
in rarer environments dominating both SMBH growth and star formation at
later cosmic times.

To first order, these results are consistent with a picture in which
energy from AGN helps heat or blow out reservoirs of gas, causing
declines in both star formation and accretion.  However, observations
show that stellar populations and AGN activity do not evolve in
lock step.  IR AGN and optically-selected Seyferts appear to be more
weakly clustered than normal galaxies matched in luminosity and color
or stellar age \citep{hickox_r_hick09corr, hickox_r_li06agnclust}, while some studies
have found that radio AGN are found in more massive dark matter halos
than a control galaxy sample \citep{hickox_r_wake08radio, hickox_r_mand09agnclust}.
This indicates that AGN fueling depends on local environment as well
as the supply of cold gas.  Further, it is not completely clear that
AGN feedback {\em causes} star formation to decline.  The X-ray AGN
content in post-starburst galaxies is higher than for a control sample
\citep{hickox_r_brow09psb}, and early-type galaxies hosting X-ray AGN have mainly ``green'' colors similar to post-starbursts (Fig.~\ref{fig:colmag}, {\em
bottom right}\citep{hickox_r_scha09agn}). However, for the \citet{hickox_r_scha09agn}
 sample, the accretion timescale is shorter than that for stellar
 evolution, indicating that the observed AGN activity actually trails
 the shutoff of star formation.  Clearly, even at relatively late
 times ($z < 0.7$) the interplay between black hole growth and star
 formation is complex and warrants a great deal of further study.

Observations at higher redshifts ($z\sim 2$--3) probe the epoch of
massive elliptical formation and peak quasar activity and allow us to
test the most powerful modes of AGN feedback invoked to quench star
formation in massive galaxies.  Recent studies indicate that many
$z\sim 2$ star-forming galaxies contain AGN
\citep{hickox_r_dadd07comp, hickox_r_fior08agn}, but statistics are limited by the relatively small areas
of the deepest X-ray fields.  There also is some evidence that highly
obscured AGN with rapid starbursts transition to a less obscured phase
with declining star formation
\citep[e.g.,][]{hickox_r_dey08dogs}, as might be expected for radiatively-dominated feedback
\citep{hickox_r_hopk08frame1}.  Whether the quenching of star formation is
caused by feedback or exhaustion of the fuel supply, and whether these
episodes are primarily triggered by cold gas-rich mergers or other
processes, remain open observational questions.

\section{New observational tests}

These and many other recent results provide motivation for more
detailed observations of the co-evolution of AGN and stellar
populations over a range of redshifts. A practical observational goal
is to derive AGN observables (such as luminosity functions and
clustering) {\em as a function of host galaxy properties}, and then
compare these directly to the output of hydrodynamic or semi-analytic
evolution models.  Such studies are particularly well-suited to X-ray
observations which, although they can miss weak or highly obscured
AGN, are the most efficient and complete tracers of accretion at
relatively high redshifts ($z > 1$).  Some recent works have begun to
study X-ray AGN clustering and luminosities in bins of host galaxy properties
\citep[e.g.,][]{hickox_r_silv08host, hickox_r_coil09xclust, hickox_r_geor09xmorph}, but these are still limited by
relatively small numbers of objects.  Large-area sensitive surveys
with {\em Chandra} and {\em XMM} would enable significant advances,
but a very large step forward will come from the next generation of
wide-field X-ray surveys.  In particular, the proposed {\em Wide-Field
X-ray Telescope}\footnotemark\ would perform large surveys to
detect $>10^7$ AGN, of which $>10^6$ would have host galaxy colors and
morphologies derived from optical and IR imaging.  The {\em WFXT} sample would
enable high-precision luminosity functions and clustering measurements
(comparable to those currently available for all X-ray AGNs) in each of
$>20$ bins of galaxy luminosity, mass, or stellar age, as well as in
4--5 bins of redshift. Thus {\em WFXT} would directly trace the
co-evolution of black holes and galaxies in unprecedented detail,
providing exciting insights into the nature of feedback.

\footnotetext{\url{http://wfxt.pha.jhu.edu/}}
 


\begin{theacknowledgments}
The author thanks the members of the Bo\"{o}tes 9 deg$^2$
multiwavelength survey collaboration.  This work is funded in part by
an SAO Postdoctoral Fellowship.
\end{theacknowledgments}






\end{document}